\documentstyle[12pt]{article}

\def\beq{\begin{equation}}
\def\eeq{\end{equation}}
\def\bea{\begin{eqnarray}}
\def\eea{\end{eqnarray}}
\def\bq{\begin{quote}}
\def\eq{\end{quote}}

\def\APP{{\it Acta Phys.Pol.} }

\def\FP{{\it Fortschr.Physik} }

\def\MPL{{\it Mod.Phys.Lett.} }

\def\NC{{\it Nuovo Cimento} }
\def\NP{{\it Nucl.Phys.} }
\def\PL{{\it Phys.Lett.} }
\def\PR{{\it Phys.Rev.} }
\def\PRL{{\it Phys.Rev.Lett.} }

\def\PTP{{\it Progr.Theor.Phys.} }
\def\PZEKF{{\it Pis'ma v Yh.Eksp.Teor.Fiz.} }

\def\ZP{{\it Z.Phys.} }

\def\gappeq{\mathrel{\rlap {\raise.5ex\hbox{$>$}}
{\lower.5ex\hbox{$\sim$}}}}

\def\lappeq{\mathrel{\rlap{\raise.5ex\hbox{$<$}}
{\lower.5ex\hbox{$\sim$}}}}

\begin{document}
\pagestyle{empty}
\begin{flushright}
{CERN-TH/95-317}
\end{flushright}
\vspace*{5mm}
\begin{center}
{\bf TESTING THE STANDARD MODEL AND BEYOND} \\
\vspace*{1cm}
{\bf John Ellis} \\
\vspace{0.3cm}
Theoretical Physics Division, CERN \\
CH - 1211 Geneva 23 \\
\vspace*{2cm}
{\bf ABSTRACT} \\ \end{center}
\vspace*{5mm}
\noindent
This paper is based on lectures presented to mathematical physicists and
attempts to
provide an overview of the present status of the Standard Model, its
experimental
tests, phenomenological and experimental motivations for going beyond
the Standard
Model via supersymmetry and grand unification, and ways to test these
ideas with particle
accelerators.

\vspace*{5cm}


\begin{flushleft}
CERN-TH/95-317 \\
November 1995
\end{flushleft}
\vfill\eject

\setcounter{page}{1}
\pagestyle{plain}

\section{Introduction to the Standard Model and its (Non-Topological)
Defects}

When we phenomenological particle physicists talk of the Standard Model,
we include
QCD, our theory of the strong interactions and the
Glashow-Weinberg-Salam electroweak
theory \cite{gws}.  Much of this lecture will be concerned with the
following fundamental
question: why are the masses of the force-carrying gauge bosons of the
Standard Model so
different, whilst their couplings to matter are so similar?
Phenomenologists believe that the
answer to this question is provided by some variant of the Higgs
mechanism, but we do not
yet have any direct experimental evidence for this belief.  However,
precision
electroweak data are beginning to provide us with some indications on
the nature of
Higgs physics, as discussed in Section 3, and may be providing us with
some
experimental motivation for supersymmetry, as discussed in Section 4.
The recent big news
in the current catalogue of the elementary particle constituents of
matter
has been the confirmation of the discovery of the top quark \cite{top},
with a
mass close to predictions \cite{predmt} based on precision electroweak
data \cite{ewdata}, as
we shall review in Section 2.
 As you probably know,
one of the first major facts established by the LEP accelerator was that
there are no
more light neutrino species. Within the Standard Model context, this
limits the number of
lepton doublets, and hence presumably means that there are no more
charged leptons either, and
(in order to cancel triangle anomalies) hence no more quarks in
generations like the three
which are now known.

The Standard Model outlined in the previous paragraph has been tested
and verified,
by experiments at LEP in particular, with a precision which is now
better than 1 \%
\cite{ewdata}. Although the Standard Model has passed (almost) all
these tests with flying
colours,  it has many (non-topological) defects which motivate going
beyond it. Theoretically,
the Standard Model is very unsatisfactory because it provides no
explanation for the
elementary particle quantum numbers (colour, electroweak isospin,
hypercharge), and
contains twenty or more arbitrary parameters. We would dearly
love to reduce the number of these parameters!

The major classes of problem that motivate going beyond the Standard
Model are three.

\noindent {\bf The Problem of Mass:} What are the origins of the
different particle
masses? Is there an elementary Higgs boson? Why are all the particle
masses so much
smaller than the Planck mass, the only candidate we have for a
fundamental scale in
physics? Does supersymmetry play a role \cite{susy} in answering this
question? As discussed in
Section 4, there are good reasons to expect that this set of questions
may be
answered by experiments performed at forthcoming accelerators, in
particular the LHC
as discussed in Section 6.

\noindent {\bf The Problem of Unification:}
Is there a simple gauge framework which includes all the interactions of
the Standard
Model? Does this yield novel phenomena such as proton decay and neutrino
masses
which can be detected, possibly by non-accelerator experiments?

\noindent{\bf The Problem of Flavour:}
Why are there so many different types of quarks and leptons? Why are the
couplings of
the $W^\pm $ mixed? What is the origin of CP violation? Some
phenomenologists suggest
that these questions may be answered in a composite model of quarks and
leptons.
Personally, I have never seen a composite model that I find convincing.
Moreover,
there is
no experimental indication on the scale at which these flavour questions
might be answered. I believe that obtaining the answer to this question
will have to wait for a
better understanding of string theory.

\noindent String theory is the only serious candidate we have for a
Theory of
Everything which includes gravity as well as the Standard Model
interactions
described above, reconciles gravity with quantum mechanics \cite{emn},
explains the origin of
the space-time, tells us why we live in four dimensions, etc. Since the
scope of
this lecture is purely phenomenological, I will not address here these
fascinating problems.

\section{Testing the Standard Model}

The electroweak sector of the Standard Model has been tested in a large
variety of
experiments at a vast range of energies and distance scales. These
extend from
measurements of parity violation in atoms \cite{pv}, with an effective
invariant momentum
transfer $Q^2$ of about 10$^{-10}$ GeV$^2$, through neutrino-electron
scattering
at $Q^2\sim 0.1$ GeV$^2$  \cite{nue}, deep-inelastic electron-, muon-
and neutrino-hadron
scattering at $Q^2 \sim 1-100$ GeV$^2$,  electron-positron collisions at
$Q^2
\lappeq 10^4$ GeV$^2$ and proton-antiproton colliders at $Q^2\sim 10^4$
GeV$^2$. The largest
momentum transfers of all have been seen in deep-inelastic
electron-proton collisions
at HERA \cite{hera}, but these are not yet of sufficient precision to
provide sensitive tests.

The most sensitive tests of the Standard Model are those provided
\cite{ewdata} by
electron-proton collisions  in the LEP accelerator at CERN, and the SLC
accelerator at SLAC.
Most of the data taken at these accelerators so far have been in the
neighbourhood of the
$Z^0$ peak \cite{lep1}, which is  perhaps the most precisely studied
Breit-Wigner peak in
history. The following are
the basic measurements performed on the $Z^0$ peak.

\noindent{\bf The Total Hadronic Cross Section:} At the tree level, this
is given by
\beq
\sigma^0_h = {12\pi\over
m^2_Z}~~{\Gamma_{ee}\Gamma_{had}\over\Gamma^2_Z}
\label{1}
\eeq
where $M_Z$ and $\Gamma_Z$ are the mass and total decay rate of the
$Z^0$ boson,
respectively, and $\Gamma_e$, $\Gamma_h$ are its partial decay rates
into
electron-positron pairs and hadrons, respectively. After including
electromagnetic
radiative corrections, the cross section in Eq. (\ref{1}) is reduced to
about
30 mb \cite{lep1}. The total event rate at LEP is given by the product
of this cross section
and the luminosity  (collision rate) which is $\lappeq 2\times 10^{31}$
cm$^{-2}$
s$^{-1}$, yielding almost one event per experiment  per second.

\noindent{\bf The Total $Z^0$ Decay Rate:} In the absence of exotic
decay modes, this
can be written in the form
\beq
\Gamma_Z = \underbrace{\Gamma_{ee}+\Gamma_{\mu\mu} +
\Gamma_{\tau\tau}}_{3\Gamma_{\ell\ell}} +
N_\nu \Gamma_\nu + \Gamma_{had}
\label{2}
\eeq
where the three leptonic decay rates are equal if one assumes
universality, and
$N_\nu$ is the number of light neutrino species. In the Standard Model,
\beq
\Gamma_\nu = 1.992 \pm 0.003 \Gamma_{\ell\ell}~.
\label{3}
\eeq
Since the neutrinos are not seen directly in the experiment, they cannot
be
distinguished from other weakly-interacting neutral particles, so the
total
\beq
\Gamma_{invisible} \equiv N_\nu \Gamma_\nu
\label{4}
\eeq
may be parametrized by a non-integer value of $N_\nu$!

\noindent{\bf Partial Decay Rates:}
By looking at particular final states, it is possible to disentangle
various partial
decay rates of the $Z^0$. Particularly accurately measured are the
$\Gamma_{\ell\ell}$, which can be related to the ratios  $R_\ell \equiv
\Gamma_{had}/\Gamma_{\ell\ell}$. Of special recent interest \cite{rbrc}
have been the partial
decay rates into bottom and charm quarks, parametrized by $R_b =
\Gamma_{b\bar
b}/\Gamma_{had}$.

\noindent{\bf Forward-Backward Asymmetries:}  At the tree level, it is
possible to
parametrize the angular distribution of $f\bar f~~~(f \not= e)$ final
states by
\beq
{d\sigma\over d\cos\theta} (e^+e^- \rightarrow \bar ff) \simeq
(1+\cos^2\theta)\cdot
F_1 + 2\cos\theta\cdot F_2
\label{5}
\eeq
One can then define the forward-backward asymmetry
\beq
A_{FB} \equiv {\int^1_0 - \int^0_{-1} \over \int^1_0 + \int^0_{-1}} =
{3\cdot F_2\over
4\cdot F_1}
\label{6}
\eeq
which has the value $3(1-4\sin^2\theta_W)^2$ for $\mu^+\mu^-$ and
$\tau^+\tau^-$.
This measurement  is particularly free of systematic detector effects
and is limited
essentially by statistics.

\noindent{\bf Final-State $\tau$ Polarization:}  The heavy lepton $\tau$
analyzes its
own polarization when it decays, which can be measured in a number of
hadronic and
leptonic final states. At the tree level in the Standard Model, the
$\tau$
polarization is given by
\beq
P_\tau = {2(1-4\sin^2\theta_W)\over 1 + (1-4\sin^2\theta_W)^2}
\label{7}
\eeq
This is a particularly sensitive way of measuring $\sin^2\theta_W$,
though again
limited by statistics.

\noindent{\bf Polarized-Beam Asymmetry:}
If a longitudinally-polarized electron beam is available, as at the SLC
\cite{slc}, one can
measure the total cross-section asymmetry
\beq
A_{LR} \equiv {\sigma_L-\sigma_R\over \sigma_L + \sigma_R} =
{2(1-4\sin^2\theta_W)\over 1 + (1-4\sin^2\theta_W)^2}
\label{8}
\eeq
where $L$ and $R$ label the different electron helicities. The electron
and positron
beams circulating at LEP have a natural transverse polarization
\cite{pt}, which is useful for
calibrating the beam energy and hence measuring the $Z$ mass and width,
as
discussed shortly, but there are no plans at CERN to rotate the beam
polarization
to the longitudinal direction.

The precision electroweak measurements from high-energy experiments at
CERN \cite{ewdata},
SLAC \cite{slc} and Fermilab \cite{fnal}, are summarized in Table 1.
Particularly notable is
the high precision ($2\times 10^{-5}$) with which the $Z$ mass is
measured. The other LEP
measurements are also considerably more precise than was thought
possible before LEP
started operation \cite{lep1}. The latest value for the total number of
neutrino species
is \cite{ewdata}:
\beq
N_\nu = 2.991 \pm 0.016
\label{9}
\eeq
I had always hoped that this number would turn out to be non-integer,
such as $\pi$
or (even better) $e$, reflecting the presence of exotic physics, but
this was not to
be.

\begin{center}
\begin{tabular}{|l|lcll|} \hline
$M_Z$ & 91.1884 &$\pm$ &0.0022 & GeV \\
$\Gamma_Z$ & ~~2.4963 &$\pm$ &0.0032& GeV \\
$\sigma^0_h$ & 41.488 &$\pm$ &0.078 &nb\\
$R_L$ & 20.788~ &$\pm$ &0.032 & \\
$A^L_{FB}$ & ~~0.0172 & $\pm$ & 0.0012 & \\
$A_\tau$ & ~~0.1418 & $\pm$ & 0.0075 &\\
$A_{\ell}$ & ~~0.1390 & $\pm$ & 0.0089 & \\
$R_b$ & ~~0.2219 & $\pm$ & 0.0017 & \\
$R_c$ & ~~0.1543 & $\pm$ & 0.0074 & \\
$A^b_{FB}$ & ~~0.0999 & $\pm$ & 0.0031 & \\
$A^c_{FB}$ & ~~0.0725 & $\pm$ & 0.0058 & \\
$\sin^2\theta_{eff}(Q_{FB})$ & ~~0.2325 & $\pm$ &0.0013 & \\
$M_W$ & ~80.26 & $\pm$& 0.16 & GeV \\
$\sin^2\theta_{eff}(A_{LR})$ & ~~0.23049 & $\pm$ & 0.00050 & \\ \hline
\end{tabular}
\end{center}
\begin{center}
- Table 1 -
\end{center}

As already mentioned, the transverse polarization of the LEP beams is
useful to
calibrate the beam energy, because the polarization disappears at
certain resonance
energies which are determined by the electron's anomalous magnetic
moment. Using this
technique, it has been possible to measure the LEP beam energy with a
precision
better than 1 MeV \cite{energy}. When this was first  done, it was
discovered that the beam
energy varied systematically by 10 MeV or more, considerably more than
the quoted error in
the $Z$ mass. Over time, these variations in the LEP beam energy have
become better
understood, and reveal many subtle and amusing effects, in addition  to
banal
effects associated with the temperature and humidity in the LEP tunnel.
For example,
as seen in Fig. 1, the energy of the LEP beam is correlated with the
positions of the
Sun and Moon \cite{pt}, which exert tidal effects on the rock in which
the LEP ring is
embedded, causing it to expand and contract, which the tuning of the
machine converts
into a variation in its energy. Even after this effect was taken into
account,
significant variations in the size of the LEP ring were detected, as
seen in Fig. 1.
Most of the variation in  1993 turned out to be correlated with the
height of the
water table inside the Jura mountains \cite{water}: water in the rock
causes it to expand,
carrying LEP with it.  However, this was not responsible for the
variations seen in
the first part of 1994. These were largely explained by the Swiss policy
of emptying
Lake Geneva in the spring, to make room for the run-off  water from the
melting snows
in the mountains. The rock surrounding LEP expands during the months
after the burden
of this water is released, much as Scandinavia is still rising after the
Ice Age.

Another bizarre effect that has been identified very recently is that of
electric trains on
the nearly railway line from Geneva to France. Not all of the
return current passes through the rails, but some passes through the
earth, and in
particular through the LEP ring, which is a relatively good conductor.
This can produce
changes in the LEP magnets corresponding to a shift of several MeV in
the beam energy. as
seen in Fig. 2  \cite{bull}. The ``TGV effect" on the LEP determination
of the $Z^0$ mass
remains to be evaluated, but seems unlikely to affect significantly the
LEP determination of
the $Z^0$ decay width \cite{jwenn}.

Figure 3 shows the implications of some of the precision measurements in
Table 1 for
the couplings of the $Z^0$ to charged leptons. We see that $g_A$ is
close to the
value -1/2 predicted in the Standard Model at the tree level, while
$g_V$ is
significantly different from zero, as expected in the Standard Model
with
$\sin^2\theta_W < 1/4$. Figure 3 also shows predictions in the Standard
Model for
different values of the top-quark and Higgs-boson
masses $m_t$ and $M_H$. As was pointed out by
Veltman \cite{tini} in particular, the precision electroweak
measurements in Table 1 are
sensitive to quantum corrections associated with unseen particles. For
example, at
the one-loop level, the $W$ and $Z$ masses are given by \cite{lep1}
\beq
m^2_W\sin^2\theta_W = m^2_Z\cos^2\theta_W \sin^2\theta_W =
{\pi\alpha\over\sqrt{2}
G_\mu}~(1+\Delta r)
\label{10}
\eeq
The radiative correction $\Delta r$
receives an important corrections from the massive top quark. If it were
absent,
the third quark isospin doublet of the Standard Model would be
incomplete, breaking
gauge symmetry and destroying the renormalizability of the Standard
Model. The
quantity $m_t^2 - m_b^2$ is a measure of electroweak isospin breaking,
which is
sensed by precision electroweak measurements through the
vacuum-polarization (oblique)
diagram shown in the first part of Fig.~4. These make a contribution
\cite{tini},  \cite{lep1}
\beq
\Delta r \ni {3 G_\mu\over 8\pi^2\sqrt{2}}~m^2_t\quad\quad{\rm
for}\quad\quad m_t \gg
m_b ~.
\label{11}
\eeq
The Higgs boson also contributes to $\Delta r$. Again, the Standard
Model would not
be renormalizable if the gauge symmetry were broken explicitly, rather
than
spontaneously. In agreement with a screening theorem proved by Veltman
\cite{tini}, the
sensitivity to the physical Higgs-boson mass provided by the last two
diagrams in Fig. 4 is
only logarithmic  \beq
\Delta r \ni {\sqrt{2} G_\mu \over 16\pi^2}~~m^2_W~~\left\{ {11\over
3}~\ln
{}~~{M^2_H\over m^2_Z} \ldots \right\} \quad\quad {\rm for}\quad\quad M_H
\gg m_W
\label{12}
\eeq
but experiments are now also sensitive \cite{efl} to the parameter
$M_H$.

Figure 5 shows  the numerical sensitivity of $\Delta r$ to $m_t$ and
$M_H$   \cite{lep1}. A
measured value of $\Delta r$ does not determine uniquely both $m_t$ and
$M_H$, since
a trade-off between their contribution is possible, but a combination of
many
different precision electroweak measurements does allow $m_t$ and $M_H$
to be
disentangled. Global fits to the precision electroweak data now use many
calculations \cite{dima} of higher-order effects going considerably
beyond (\ref{10}),
(\ref{11}), (\ref{12}).

Combining all the available precision electroweak data from LEP, SLC,
Fermilab and
low-energy $\nu q$, $eq$, $\mu q$ and $\nu e$ interactions, a  global
fit with $M_H$
left as a free parameter predicts \cite{efl}
\beq
m_t = 155 \pm 14 ~{\rm GeV}
\label{13}
\eeq
as seen in Fig. 6. The contributions of the different sectors to the
$\chi^2$ function are
shown in Fig. 7. A somewhat higher value is obtained if the LEP data
alone are used in the fit, and the central value of $m_t$ is increased
substantially  if $M_H$ is not left
free, but is fixed at 300 GeV \cite{ewdata}. The indirect determination
(\ref{13}) is
consistent (within errors) with the mean value of the published CDF and
D0 measurements
\cite{top}:  \beq
m_t = 181 \pm 12~{\rm GeV}~.
\label{14}
\eeq
It is therefore appropriate to make a combined fit of the direct and
indirect
measurements (which have comparable weights), yielding \cite{efl}
\beq
m_t = 172 \pm 10~{\rm GeV}~.
\label{15}
\eeq
With the recent discovery of the top quark at Fermilab, the ``Mendeleev
table" of
elementary matter constituents is apparently now complete. Now the fun
starts, namely
solving the problem of mass and finding the Higgs boson, or whatever
replaces it.

\section{The Electroweak Vacuum}

It is generally accepted by theorists that generating the masses of the
particles in
the Standard Model requires a spontaneous breakdown of its gauge
symmetry
\beq
m_{W,Z} \not= 0 \Leftrightarrow < 0\vert X_{I,I_3}\vert o> \not= 0
\label{16}
\eeq
where $X$ is some field with non-trivial isospin and a non-zero vacuum
expectation
value. Measurements of the $W$ and $Z$ masses
\beq
\rho = {m^2_W\over m^2_Z \cos^2\theta_W} \simeq 1
\label{17}
\eeq
indicate that the field $X$ mainly has isospin $I = 1/2$  \cite{rv}.
This is also what is
required to give masses to the quarks and leptons in the Standard Model:
\beq
\lambda_f ~H_{I=1/2} \bar f_Lf_R \Rightarrow m_f ~\bar f_L~f_R
\label{18}
\eeq
There is a general consensus on the above statements: however, you can
start an
argument when you discuss whether $X$ is elementary or composite.

The option chosen in the original formulation of the Standard Model by
Weinberg and
Salam \cite{gws} was that of an elementary Higgs boson: $<0\vert
H^0\vert 0> \not= 0$. This is
fine at the classical tree level, but yields problems when you calculate
quantum
loops. Each individual one of the diagrams shown in Fig. 8 yields a
quantum
correction
\beq
\delta M^2_H \simeq 0\left({\alpha\over\pi}\right) \Lambda^2
\label{19}
\eeq
where $\Lambda$ represents a cut-off in momentum space, above which the
Standard
Model is modified or replaced. As discussed in the next section, these
quantum
corrections may be reduced to $\lappeq m^2_H$ if one invokes \cite{susy}
supersymmetry at an
energy scale below 1 TeV.

The alternative option is to postulate that $X$ is composite, presumably
a condensate
of strongly-interacting fermion-antifermion pairs:
\beq
< 0\vert \bar FF\vert 0> \not= 0
\label{20}
\eeq
by analogy
with quark condensation in QCD, and the condensation
of Cooper pairs in conventional superconductivity. Possible candidates
for the
strongly-interacting fermion $F$ include the top quark \cite{tcond} ,
which could be bound by
strong Yukawa couplings if it were sufficiently heavy, or techniquarks
$T$  \cite{tc} bound by
new technicolour interactions at an energy scale of order 1~TeV.

The precision electroweak data reviewed in the previous section already
provide us
some indications on $M_H$  \cite{efl}, which seem to disfavour the
available composite Higgs
scenarios. The correlation between $m_t$ and $M_H$ seen in Fig. 5 is
weakened when
one makes a global fit to all the high- and low-energy data, as seen in
Fig. 9.
Indeed, as seen in Fig. 10, a global fit provides a $\chi^2$ function
which looks
Gaussian as a function of log $M_H$, even before the direct CDF and $D0$
measurements
of $m_t$ are included. As also seen in Fig. 9, the data prefer a
relatively light
Higgs boson. If we do not include the direct measurements of $m_t$, we
find \cite{efl}
\beq
M_H = 36^{+56}_{-22}~{\rm GeV}
\label{21}
\eeq
which becomes
\beq
M_H = 76^{+152}_{-50}~{\rm GeV}
\label{22}
\eeq
if the direct Fermilab
measurements \cite{top} are included.  The range in Eq. (\ref{22}) can
be rephrased as
\beq
\log_{10} ~~\left({M_H\over M_Z}\right) = -0.08^{+0.48}_{-0.46}
\label{23}
\eeq
which is perhaps more appropriate in view of the logarithmic sensitivity
to $M_H$. As
seen in Fig. 11, this preference for a relatively light Higgs boson has
been a
consistent trend for several years \cite{oldefl}. Moreover, it is now
confirmed by several
other recent global fits to the available electroweak data
\cite{ewdata},  \cite{others}. As
discussed in more detail in the next section, the preferred value of
$M_H$ is highly
consistent with the range predicted in the minimal supersymmetric
extension of the Standard
Model (MSSM). Independently of this theoretical prejudice, I now offer
3-to-1 odds that $M_H
<$ 300 GeV.

The relatively low value  Eq. (\ref{21}) and (\ref{22}) bodes ill for a
composite
Higgs model. Minimal scenarios for $t\bar t$ condensation based on a
Nambu-Jona
Lasinio model
\beq
{\cal L}_{NJL} = \bar\psi_a D\llap{$/$} \psi^a + {1\over 2} ~G~\left[
(\bar\psi_a\psi^a)^2 - (\bar\psi_a \gamma^5\psi_a)^2\right]
\label{24}
\eeq
correspond to a reformulation of the Standard Model with constraints
that lead to
\cite{tcond}  \beq
M_H \simeq (1~{\rm to}~2) m_t~~:~~m_t \sim 200~{\rm to}~ 250~{\rm GeV}
\label{25}
\eeq
Neither of these  predictions agrees well with experiment, in particular
the
top quark appears to be too light. This has not completely discouraged
would-be
top-quark condensers, some of whom are postulating epicycles such as
supersymmetry
and/or an extension of the Standard Model gauge group \cite{newtcond}.

Technicolour \cite{tc} would be able to provide masses for the $W$ and
$Z$ with just one
isospin doublet of techniquarks $T$, whose new gauge interactions would
become
strong at an energy scale
\beq
\Lambda_{TC} \simeq 3000 \Lambda_{QCD}
\label{26}
\eeq
However, this minimal model requires some extension if it is to provide
fermion
masses, and the conventional scenario \cite{etc} discussed is a model
with one
technigeneration: \bea
(\nu , e)\phantom{xxxxxxx} & (u, d)\phantom{xxxxxxx} \nonumber \\
(N, E)_{1,\ldots , N_{TC}} & (U, D)_{1,\ldots , N_{TC}}
\label{27}
\eea
This model has long had potential problems with light charged
technipions and a
possible flavour-changing neutral interactions, which have motivated
variants such as
``walking" technicolour \cite{walk}. The miseries of this model have
been compounded by recent
precision electroweak data.

The quantum effects of a large class of extensions of the Standard Model
which add new
isospin representations, including the above-mentioned technicolour
model, can
largely be characterized by their effects on three combinations of
bosonic vacuum
polarizations \cite{stu}:
\bea
T &\equiv & {\epsilon_1\over\alpha} \equiv {\Delta\rho\over\alpha}
\nonumber \\
\Delta\rho &= & {\pi_{XX}(0) \over m^2_Z} - {\pi_{WW}(0) \over m^2_W} -
\tan\theta_W~~{\pi_{\gamma Z}(0) \over m^2_Z} \nonumber \\
S &\equiv & {4\sin^2\theta_W\over\alpha} ~\epsilon_3~,\quad U \equiv
-{4\sin^2\theta_W\over\alpha}~\epsilon_2
\label{28}
\eea
in the Standard Model, the leading behaviours of $T$ and $S$
 are
\bea
T &=& {3\over 16\pi}~~{1\over \sin^2\theta_W \cos^2_W}~~{m^2_t\over
m^2_Z} - {3\over 16\pi
\cos^2\theta_W}~~\ln~\left({M^2_H\over M^2_Z}\right) + \ldots \nonumber
\\
S &=& {1\over 12\pi}~\ln \left({M^2_H\over M^2_Z}\right) +  \ldots
\hfill
\label{29}
\eea
as functions of $m_t$ and $M_H$. The previous constraints on $m_t$ and
$M_H$ may be
regarded, alternatively, as bounds on new physics contributions to $S,
T, U$
\cite{stu},  \cite{abc}. Figure 12 shows as an example one analysis of
the constraints on
these variables found \cite{efltc} in a global fit, in which a fourth
parameter $\epsilon_b$ is
introduced to parametrize quantum corrections to the $Z b\bar b$ vertex
\cite{abc}. Also shown
in Fig. 12 are the Standard Model predictions, shown as a grid for
different values of $m_t$
and $M_H$, and the range of possible predictions in a minimal one-family
technicolour model
with $N_{TC} = 2$. The technicolour model is apparently disfavoured, but
some
possible modifications of its predictions could be envisaged
\cite{efltc}, as indicated by the
arrows in Fig. 12. Discarding for the moment these possibilities, and
disregarding
the uncalculable possibility of ``walking" technicolour \cite{walk},
Fig. 13 shows the price
that one must pay in order to reconcile a minimal technicolour model
with
precision electroweak data.

It seems that the Higgs boson is likely to be relatively light, in
apparent conflict
with the available strongly-interacting models. The indications on $M_H$
presently
available are likely to become strengthened during the coming decade
\cite{dpf}, as seen in
Fig. 14. We may even discover the Higgs! As seen in Fig. 15, the LEP2
accelerator now
starting to provide data should enable us to explore Higgs masses up to
about 95 GeV
\cite{lep2}. This already covers much of the range favoured by the
present data shown in Fig.
9, and also explores much of the MSSM parameter space, as discussed in
the next section.

\section{Motivations for Supersymmetry}

Supersymmetry \cite{oldsusy} is a  beautiful theory,
but the motivations for it to appear at accessible energies are related
to the problem
of mass mentioned above, namely the origin of the hierarchy of mass
scales in physics,
and its naturalness in the presence of radiative corrections
\cite{susy}. The question why
$m_W$ is much less than $m_{\rm Planck}$ or $m_{\rm GUT}$ can be
rephrased as the question:
why is $G_F \lappeq G_N$, or even why the Coulomb potential inside an
atom is much stronger
than the Newtonian potential:
\beq
{e^2 \over r} \lappeq G_N \times {m^2 \over r}
\label{30}
\eeq
This hierarchy is valuable to radiative corrections. We say that a
theory is natural if
the radiative corrections are not much larger than the physical values
of observable
quantities. For example, the leading one-loop correction to a fermion
mass takes the
form
\beq
\delta m_f = 0\left({\alpha\over\pi}\right)~m_f~~\ln \left({\Lambda\over
m_f}\right)
\label{31}
\eeq
which is not much larger than $m_f$ for any reasonable cut-off $\Lambda
\gappeq  m_P$.

Naturalness is, however, a problem for an elementary Higgs boson, which
in the
electroweak sector of the SM must have a mass
\beq
m_H = m_W \times 0\left(\sqrt{{\alpha\over\pi}}\right)^{0\pm 1}~.
\label{32}
\eeq
As already mentioned, the one-loop diagrams shown in Fig. 8 lead to
``large" radiative
corrections of the form
\beq
\delta m^2_{H,W}\simeq 0\left({\alpha\over\pi}\right)~~\Lambda^2~.
\label{33}
\eeq
These are much larger than the physical value $m_H^2$ for a cut-off
$\Lambda$,
representing the scale at which new physics appears, of order $m_P$ or
$m_{\rm GUT}$.

Supersymmetry solves the naturalness problem of an elementary Higgs
boson by
virtue of the fact that it has no quadratic divergences and fewer
logarithmic
divergences than non-supersymmetric theories. The diagrams shown in Fig.
8 have
opposite signs, so that their net result is  \beq
\delta m^2_{W,H} \simeq - \left({g^2_F\over 4\pi^2}\right)~~(\Lambda^2 +
m^2_F) +
\left( {g^2_B\over 4\pi^2}\right) ~~(\Lambda^2 + m^2_B)~.
\label{34}
\eeq
The leading divergences cancel if there are the same numbers of bosons
and
fermions, and if they have the same couplings $g_F = g_B$, as in a
supersymmetric theory. The residual contribution is small if
supersymmetry is
approximately valid, i.e., if $m_B \simeq m_F$:
\beq
\delta m^2_{W,H} \simeq 0 \left({\alpha\over\pi}\right)~~\left\vert
m^2_B -
m^2_F\right\vert
\label{35}
\eeq
which is no larger than $m_{W,H}^2$ if
\beq
\left\vert m^2_B - m^2_F\right\vert \lappeq 1~{\rm TeV}^2
\label{36}
\eeq
This property provides the first motivation for supersymmetry at low
energies.
However, it must be emphasized that this is a qualitative argument which
should be regarded as a matter of taste. After all, an unnatural theory
is
still renormalizable, even if it requires fine tuning of parameters. A
second
supersymmetric miracle is the absence of many logarithmic divergences:
for many
Yukawa couplings and quartic terms in the effective potential,
\beq
\delta\lambda \propto \lambda
\label{37}
\eeq
which vanishes if the rare coupling $\lambda = 0$. The combination of
Eqs.
(\ref{35}) and (\ref{37}) means that if $M_W \leq M_P$ at the tree
level, it stays
small in all orders of perturbation theory, solving the naturalness
problem and
providing a context for attacking the hierarchy problem \cite{susy}.

The latter is particularly accute in theories with both large and small
scales, in which the
former may ``leak" and contaminate the latter \cite{gildener}. Consider
for example a Grand
Unified Theory with two sets of Higgs bosons, $H$ with a large vacuum
expectation value
$V_{GUT}$ and $h$ with a small vacuum expectation value $v_{EW}$. In a
generic Grand Unified
Theory, there will be a quartic coupling $\lambda hhHH$, which yields
\beq
\delta m^2_H \simeq \lambda \cdot V^2_{\rm GUT}
\label{38}
\eeq
which is a large and potentially disastrous contribution to the light
Higgs mass. Even if
$\lambda = 0$ at the tree level (why? this is the hierarchy problem),
radiative corrections
will regenerate a non-zero coupling, so that
\beq
\delta m^2_H \simeq 0\left({\alpha\over\pi}\right)^2~ V^2_{\rm GUT}
\label{39}
\eeq
Such contributions need to be suppressed to many orders of perturbation
theory, which
requires a powerful symmetry, such as supersymmetry. It is also worth
pointing out that it
has been argued \cite{hawk} that quantum gravity effects may also
generate a large shift in the
mass of an elementary Higgs boson
\beq
\delta m^2_H = 0(m^2_P)
\label{40}
\eeq
although to be sure of this, one needs a consistent quantum theory of
gravity. Effects such
as (\ref{40}) are likely to be absent in a supersymmetric theory, and,
in any case, the only
consistent quantum theory of gravity which we possess is string theory,
which is difficult
or impossible to formulate consistently without supersymmetry.

\section{Model Building}

Now that we are motivated to construct a supersymmetric model, the first
question is whether
the known fermions ($q, \ell$) could be the supersymmetric partners of
the ``known" bosons
($\gamma , W, Z$, $H, g$)? As was first pointed out by Fayet
\cite{fayet}, the answer is not
for phenomenology, since their quantum numbers do not match. For
example, the quarks $q$ appear
in {\bf 3} representations  of $SU(3)_c$, whereas the bosons appear in
{\bf 1} and {\bf 8}
representations. Likewise, the leptons $\ell$ have non-zero lepton
number, whereas all the
bosons have zero lepton number. As a result, one must introduce
supersymmetric partners for
all the known particles, as shown in Table 2. You may not appreciate the
economy in particles,
but you should appreciate the economy of the supersymmetric principle.

\begin{center}
\begin{tabular}{|lclc|} \hline
 &J & sparticle &J \\
$q_{L,R}$ & 1/2 & $\tilde q_{LR}$ squark & 0 \\
$\ell_{L,R}$ & 1/2 & $\tilde \ell_{L,R}$ slepton & 0 \\
$\gamma$ & 1 & $\tilde\gamma$ photino & 1/2 \\
$Z$ & 1 &$\tilde Z$ zino & 1/2 \\
$W^\pm$ & 1 &$\tilde W^\pm$ wino & 1/2 \\
$H^{\pm ,0}$ & 0 & $ \tilde H^{\pm , 0}$ higgsino & 1/2 \\ \hline
\end{tabular}
\end{center}
\begin{center}
- Table 2 -
\end{center}

You may wonder whether, if $N = 1$ supersymmetry is good, perhaps $N >
1$ supersymmetry is
better?  The answer is: not for phenomenology, because such a theory
cannot accommodate chiral
fermions. The available $N = 2$ supermultiplets are
\beq
\left(\matrix{~~{1\over 2} \cr 0,0 \cr -{1\over 2}}\right)  ~~
\left(\matrix{+1\cr
+{1\over 2} \cr 0}\right) \bigoplus \left(\matrix{0 \cr -{1\over 2} \cr
-1}\right)
\label{41}
\eeq
in which the fermions of helicity $+1/2$ have the same internal quantum
numbers as the
fermions of helicity $-1/2$, making it impossible to accommodate the
parity violation seen in
the electroweak interactions.

The starting point for any discussion of supersymmetric phenomenology is
the minimal
supersymmetric extension of the Standard Model \cite{mssm}, which has
the same gauge
interactions as the Standard Model, and whose Yukawa interactions are
derived from the
following superpotential, which is written as a holomorphic function of
left-handed
superfields: \bea
W &=& \sum_{L,E^c} \lambda_L~ LE^cH_1 + \sum_{Q,U^c} \lambda_U~ QU^cH_2
\nonumber \\
&& + \sum_{Q,D^c} \lambda_C ~QD^cH_1 + \mu H_1H_2
\label{42}
\eea
Here $L$ and $Q$ denote left-handed quark and lepton doublets,
respectively and $E^c$, $U^c$,
$D^c$ denote the conjugate lepton and quark singlets.
The first term in Eq. (\ref{42}) provides masses for the charged
leptons:
\beq
m_L = \lambda_L v_1
\label{43}
\eeq
and the next two provide masses for the charged 2/3 and charged -1/3
quarks,
respectively:
\beq
m_u = \lambda_u v_2~,\quad\quad m_d = \lambda_d v_1
\label{44}
\eeq
Notice that two Higgs doublets $H_{1,2}$ are needed in order to preserve
the holomorphy of the
superpotential $W$, and to cancel out axial $U(1)$ current anomalies,
and that the fourth
term in $W$ accommodates mixing between the two Higgs doublets.  The
Yukawa and gauge
couplings in the MSSM make a supersymmetric contribution to the
effective potential of the
form:
\beq
V = \sum_i\vert F_i\vert^2 + {1\over 2} \sum_a(D^a)^2
\label{45}
\eeq
where
\beq
F^*_i = {\partial W\over \partial\phi^i}~,\quad\quad D_a = g_a\phi^*_i
(T^a)^i_0 \phi^j
\label{46}
\eeq
are the conventional $F$ and $D$ terms, respectively. The fact that the
quartic terms in the
effective potential are so constrained provides  restrictions on the
supersymmetric Higgs
boson masses, as will be discussed later.

As you may well imagine, all the searches for supersymmetric particles
have been unsuccessful
so far, and have provided the following approximate lower limits on some
of their masses
\beq
m_{\tilde \ell, \tilde\mu , \tilde\tau , \tilde W} \gappeq 45~{\rm
GeV}~\cite{lepex},
\quad\quad m_{\tilde q, \tilde g} \lappeq 150~{\rm GeV}~~  \cite{fnal}
\label{47}
\eeq
The LEP2 energy upgrade will provide us with access to a new range of
sparticle masses, and
continuation of the Fermilab proton-antiproton collider will increase
the search range for
squarks and gluinos. Why do we phenomenologists keep the faith that
supersymmetric particles
will eventually be found, despite the lack of direct experimental
evidence?

In addition to the theoretical motivations for supersymmetry, there are
two tentative and
indirect experimental motivations provided by the precision electroweak
data, which come
mainly from LEP. One is provided
by the previously-mentioned indication that the Higgs boson is
``probably" light: $M_H
\lappeq$ 300 GeV, which is consistent with the MSSM expectation that
\cite{susyhmass}
\beq
m_h \simeq m_Z \pm 40~{\rm GeV}
\label{48}
\eeq
As mentioned above, the MSSM contains two Higgs doublets
\beq
H_2 = \left(\matrix{H^+_2\cr H^0_2}\right)~,\quad H_1 =
\left(\matrix{H^0_1\cr H^-_1}\right)
\label{49}
\eeq
which contain a total of eight real degrees of freedom. Three of these
are ``eaten" by the
$W^\pm$ and the $Z^0$ to yield their masses, leaving five physical Higgs
bosons to be
discovered by experiment.  Three of these are neutral, the scalars $h,
H$ and the superscalar
$A$, and two are charged $H^\pm$. At the tree level, all the masses
and couplings of these Higgses are controlled by two parameters, which
may be taken as ($m_A,
\tan\beta \equiv v_2/v_1$):
\bea
m^2_h + m^2_H &=& m^2_A + m^2_Z \nonumber \\
m^2_{H^\pm} &=& m^2_A + m^2_{W^\pm} \nonumber \\
m^2_{h,H} &=& {1\over 2} \left[ m^2_A + m^2_Z \mp
\sqrt{(m^2_A + m^2_Z)^2 - 4m^2_Z m^2_A \cos^22\beta}\right]
\label{50}
\eea
In particular, the lighter scalar Higgs $h$ was guaranteed to be lighter
than the $Z$, which
was good news for LEP2. However, radiative corrections associated in
particular with the
heavy top quark \cite{deltamh}
\beq
\delta m^2_h \propto {m^4_t\over m^2_W}~~\ln \left({m^2_q \over
m^2_t}\right)
\label{51}
\eeq
increase the upper limit to
\beq
m_h \gappeq 130~{\rm GeV}
\label{52}
\eeq
As seen in Fig. 16, it is still true that much of the $h$ mass range
will be explored at LEP2
\cite{lep2}, but, alas, not all of it.

As already mentioned, the range (\ref{52}) is highly consistent with the
indirect indications
from the precision electroweak data  on the possible mass of the Higgs.
One can even go
further, and argue that the LEP data slightly favour the MSSM over the
Standard
Model \cite{efl}. In the latter, the requirement that all the couplings
remain finite in the
energy range  $E \lappeq \Lambda_P$ impose an upper limit on the Higgs
mass. If the Standard
Model is to remain valid all the way up to $\Lambda_P \sim m_{GUT}$ or
$M_P$, then $M_H\lappeq
$ 200 GeV as seen in Fig. 17  \cite{upper}. On the other hand, the
(meta)stability of the
electroweak vacuum imposes a lower limit which depends on the scale
$\Lambda_V$ up to which
the effective Higgs potential is assumed to be reliable, as also seen in
Fig. 17
\cite{lower}. Thus, the Standard Model as we know it is consistent with
only a small range of
Higgs masses \beq
116~{\rm GeV} \lappeq M_H \lappeq 190~{\rm GeV}
\label{53}
\eeq
for $\Lambda_P = \Lambda_V = 10^{19}$ GeV and $m_t \simeq $ 172 GeV as
found in the previous
global fit. This is to be contrasted with the range
\beq
50~{\rm GeV} \lappeq m_h \lappeq 124~{\rm GeV}
\label{54}
\eeq
allowed in the MSSM for the same value of $m_t$. According to Fig. 18,
which is deduced from
the $\chi^2$ function in Fig. 9, the apparent probabilities of these
mass ranges are about 18
and 36 \%, respectively \cite{efl}. Therefore I offer another bet: I
offer 2-to-1 odds on the
MSSM!

The second indirect indication in favour of supersymmetry is provided by
the well-publicized
consistency of the measurements at the Standard Model gauge couplings
$\alpha_{1,2,3}$ with
the predictions of minimal supersymmetric GUTs \cite{susygut}. Ever
since
1987 \cite{amaldi},  \cite{costa}, but with a statistical strength which
has increased greatly
with the advent of LEP data, the prediction for  $\sin^2\theta_W$ in a
minimal
non-supersymmetric GUT \cite{marciano}:
\bea
\sin^2\theta_W(m_Z)\bigg\vert_{\overline{\rm MS}} &=& 0.208 + 0.004
(N_H-1) + 0.006
\ln \left({400~{\rm MeV}\over \Lambda_{\overline{\rm MS}}(N_f =
4)}\right) \nonumber \\
\nonumber \\
&=& 0.214 \pm 0.004 ~{\rm for}~ \Lambda_{\overline{\rm MS}}(4) = 200
{}~{\rm to}~ 800~{\rm MeV}
\label{55}
\eea
has been in conflict with data, which now indicate \cite{sin2}
\beq
\sin^2\theta_W(m_Z)\bigg\vert_{\overline{\rm MS}} = 0.2312 \pm 0.0003
\label{56}
\eeq
The prediction for $\sin^2\theta_W$ is less precise, even in the minimal
supersymmetric GUT,
because it contains more parameters, and it is not possible at present
to use this
consistency to provide meaningful constraints on the possible masses of
supersymmetric
particles \cite{caveat}.  Nevertheless, we are encouraged to believe
that supersymmetry may lie
``just around the corner", which means either at LEP2 or at the LHC, as
we now discuss.

\section{Physics with the LHC}

This accelerator \cite{lhc} provides us with our best prospect for
exploring the 1 TeV energy
region, where we may expect to find the Higgs boson and supersymmetry.
The LHC offers several
possibilities for colliding different types of particle. Of most
interest for new particle
searches is its proton-proton collider mode, which will have a
centre-of-mass energy of up to
14 TeV, and a luminosity of up to 10$^{34}$ cm$^{-2}$ sec$^{-1}$. Also
possible are heavy-ion
collisions with nuclei up to lead: used as a lead-lead collider, the LHC
would have a
centre-of-mass energy up to 1.2 PeV and a luminosity of up to 10$^{27}$
cm$^{-2}$ sec$^{-1}$,
whilst the luminosity could be higher if lighter calcium is used. It
will
also be possible to use the LHC as an electron-proton, proton-nucleus or
electron-nucleus
collider, if the mood so takes us. The LHC was approved by the CERN
Council at the end of
1994, to start doing physics in the 2004. For reasons of cash flow, the
initial approval was
for a machine with fewer magnets, able to reach a centre-of-mass energy
of 10 TeV to start
with. However, if non-member states contribute significantly, it may be
possible to start
immediately at the full design energy of 14 TeV: the final machine
schedule and energy will
be decided at a review in 1997.

The initial LHC experimental programme is expected to include the
following four experiments:
ATLAS \cite{atlas} and CMS \cite{cms}, which are large general-purpose
experiments for
discovery physics in proton-proton collisions, ALICE \cite{alice}, which
is primarily intended
for heavy-ion experiments searching for the quark-gluon plasma, though
it may also be used to
look for diffractive scattering, and LHC-B \cite{lhcb}, an experiment
designed primarily to
look for CP violation in the decays of B mesons produced in
proton-proton collisions.

The LHC accelerator will benefit fully from the existing CERN
infrastructure, since it will
be built inside the existing LEP tunnel, and will receive particles
which have been
pre-accelerated by  the other CERN accelerators.The LHC magnets are of a
very ambitious
design, with a high magnetic field above 9 Tesla and two magnetic
channels carrying beams
circulating in opposite directions. Successful tests have been made with
the first magnetic
prototypes, indicating that the maximum design energy should be
reachable, and may even be
exceeded. The ALICE and LHC-B experiments will be placed in underground
pits which have
already been dug for two of the LEP experiments, but the ATLAS and CMS
experiments will
require two very large new pits. These and tunnels for transferring the
proton and heavy-ion
beams from the lower-energy SPS accelerator are the main pieces of civil
engineering that
will be required. It just so happens that one of the beam transfer lines
points in the
direction of Italy and Greece, where neutrino detectors are now being
built that could be
used for long baseline neutrino experiments \cite{lbl}, using neutrinos
produced by a CERN
proton beam, but it has not yet been decided whether this will be
included in the LHC
programme.

Top of the LHC physics agenda will be the search for the Higgs boson,
which should have a
mass below about 1 TeV, indeed below about 300 GeV if one believes the
indirect indications
from precision electroweak experiments \cite{efl}, even 90 $\pm$ 40  GeV
if the
MSSM is correct \cite{deltamh}. A Standard Model Higgs boson will be
detectable at LEP2 as soon
as the centre-of-mass energy is increased a few GeV above $M_Z + M_H$
\cite{lep2}. Since the
maximum LEP2 centre-of-mass energy is expected to be about 192 GeV, this
means that a Higgs
weighing more than about 95 GeV will be prey for the LHC. The Standard
Model Higgs boson will
be detectable at the LHC by its decay into $\gamma\gamma$ if $M_H
\lappeq $ 140 GeV, by its
decay into $4 \ell^\pm$ if 130 GeV $\lappeq M_H \lappeq $ 700 GeV, and
by its
decay into $\ell^+\ell^-\bar\nu\nu$ if 700 GeV $\lappeq M_H \lappeq$ 1
TeV. Figure 19
summarizes the Higgs discovery significance that is expected by
combining the ATLAS and CMS
experiments \cite{atlas},  \cite{cms}. The vertical axis is the number
of signal events $S$
divided by the statistical fluctuation in background $\sqrt{B}$.
Discovery can be claimed if
$S/\sqrt{B} > 5$, which is seen from Fig. 19 to be the case for the full
range 90 GeV
$\lappeq M_H \lappeq $ 1 TeV.

The search for one or more MSSM Higgs bosons is more complicated,
because the product of the
production cross-section and the observable decay branching ratio is
often smaller than in
the Standard Model, and there are several Higgs bosons to be found with
a number of different signatures. Figure 20 summarizes incomprehensibly
the overall prospects
for the supersymmetric Higgs search at the LHC \cite{atlas},
\cite{cms},  \cite{daniel}.
Regions on the shaded side of each solid line can be explored by the
LHC. Diligent examination
of the two-dimensional parameter space will not reveal any region where
discovery is
impossible. We therefore conclude that the LHC will be able to prove or
disprove the MSSM via
its Higgs sector alone.

Next on the LHC physics agenda will be the search for supersymmetric
particles, which may
well turn out to be the biggest
banana of all. Figure 21 exhibits the expected cross-sections for
producing pairs of gluinos
and/or squarks at the LHC. These are expected to|have a high probability
for decays with
large missing transverse energy carried away by the lightest
supersymmetric particle, which
is expected to be a weakly-interacting neutral particle analogous to the
neutrino, but
heavier. This missing transverse-energy signature is expected to be
observable even if the
squarks and gluinos decay in a cascade through various intermediate
states before arriving at
the lightest supersymmetric particle. Figure 22 shows that the
supersymmetric signal is
expected to stick out above the irreducible total Standard Model
background and the
background due to experimental imperfections, if the gluino and squark
each weigh 1.5 TeV and
one looks for events with more than 300 GeV of missing transverse
energy. The
ATLAS \cite{atlas} and CMS \cite{cms} collaborations have concluded that
they should be able to
detect squarks and gluinos weighing anything up to about 2 TeV, which
includes all the range
motivated by the naturalness and hierarchy arguments presented
previously.

We conclude optimistically that, within ten years or so, experiments at
the LHC will be able to
confirm or refute supersymmetry, if this has not already been done by
LEP2 and/or the
Fermilab proton-antiproton collider. Therefore we may soon know the
answer to the question
whether supersymmetry is just a beautiful holomorphic theory, or also a
part of physics.

\vfill\eject

\noindent
{\bf Figure Captions}

\begin{itemize}
\item[Fig. ~1] Sensitivity of the LEP beam energy to (a) tides
\cite{energy}: the solid lines
are due to a tidal model, (b) the water table in the Jura mountains and
(c) the level of Lake
Geneva \cite{water}.

\item[Fig. ~2] (a) The ``TGV effect" on the LEP beam energy
\cite{bull}, due (b) to the
vagabond current from electric trains returning via the LEP ring.

\item[Fig. ~3]  The values of the vector and axial couplings of leptons
$(g_V, g_A)$, as
extracted from LEP data \cite{ewdata}.

\item[Fig. ~4]
Vacuum-polarization (oblique) diagrams contributing to the one-loop
radiative
corrections.

\item[Fig. ~5] The numerical sensitivity of $\Delta r$ (10) to $m_t$
(11) and $M_H$ (12). A
determination of $\Delta r$ alone cannot fix both $m_t$ and $M_H$.

\item[Fig. ~6] The dependence of the $\chi^2$ function on $m_t$ for
various assumed values of
$M_H$, from a recent global fit \cite{efl}.

\item[Fig. ~7] The contributions of the different sectors of precision
electroweak data to
the $\chi^2$ function shown in Fig. 6  \cite{efl}.

\item[Fig. ~8] Quadratically-divergent contributions to $\delta M^2_H$.

\item[Fig. ~9] Combined fit to all precision electroweak data
                 in the $(M_H,\,m_t)$ plane, including (solid lines)
		               or not (dashed lines) the direct determination
		               of $m_t$ by CDF/D0 (error bar on the left) \cite{top}.
		               The contours correspond to $\Delta\chi^2=1,\,4$
		               around the minimum (small circle) in either case.
 	             	 Notice that $M_H$ is significantly below 300 GeV
                at the $1\sigma$ level, and below 1 TeV at the
$2\sigma$ level \cite{efl} .

\item[Fig. 10] The values of $\chi^2$ as functions of $M_H$ for the
various
                 indicated values of $m_t$ \cite{efl}.

\item[Fig. 11] $\Delta \chi^2 = 1$ ranges for $m_H$ in a series of
global fits to the
available precision electroweak data \cite{oldefl}.

\item[Fig. 12] Comparison of the Born approximation (stars), projections
of the
$\Delta\chi^2 = 1,4$ ellipsoid (solid ellipses), the SM (grid) and the
predictions of a
one-generation TC model with $N_{TC} = 2$, a Dirac technineutrino, $M_U
= M_D$, 100 GeV $<
M_E <$  600 GeV, 50 GeV $< M_N < M_E $ and  the technicolour parameter
\cite{tc}   $\xi >$ 1/2
(scattered dots). The TC predictions are added to the SM radiative
corrections, using the
reference values $m_t$ = 170 GeV and $M_H = M_Z$. Noted that the TC
predictions are further
than the SM from the experimental data. The bold arrows labelled TQ and
B indicate possible
shifts in the TC predictions of definite sign, and the other (thin)
arrows labelled B and NC
indicate shifts that are less certain.

\item[Fig. 13] Contours of $\sigma\equiv \sqrt{\Delta\chi^2}$ for
one-generation models with
either Dirac technineutrinos (a), (b) or Majorana technineutrino (c),
(d). Note that $\sigma
\gappeq$ 4.5 in all of the TC parameter space, to be compared with
$\sigma$ = 2.6 in the SM
at the reference point $(m_t$ = 170 GeV, $M_H = M_Z$). In the case of
techniquark mass
degeneracy $(M_U = M_D)$, the Dirac and Majorana models fits are
comparable; in the case $M_U
> M_D$, however, the Dirac model becomes highly disfavoured. In all
cases, $\xi$ = 1/2 is
assumed.

\item[Fig. 14] Possible improvements in the precision with which $M_H$
can be estimated by
global fits to future precision electroweak data \cite{dpf}.

\item[Fig. 15] The range of $M_H$ accessible to LEP2 with a
centre-of-mass energy of 192 GeV
 \cite{lep2}.

\item[Fig. 16] Reach for Higgs bosons in the MSSM at LEP2 with a
centre-of-mass energy of
192 GeV. The dark shaded regions are excluded theoretically \cite{lep2}.

\item[Fig. 17] Comparison of combined top-Higgs mass fits in the
                 Standard Model (SM, upper plot) and in its Minimal
                 Supersymmetric extension (MSSM, lower plot), at
		               $\Delta\chi^2=1$. The continuation of the $\Delta\chi^2
= 1$
                 contour below the LEP direct
                limit $M_H>65$ GeV  \cite{ewdata} is shown dashed. Also
shown
	              	 in the SM plot are the lower limits on $M_H$ from
                 vacuum metastability  \cite{lower}
                 as a function of the ``new physics''
	                scale $\Lambda_V=10^4$--$10^{19}$ GeV \cite{upper},
                and the upper limts that come from requiring the SM
couplings
                to remain perturbative up to a scale
$\Lambda_P=10^3$--$10^{19}$ GeV.
                 In the MSSM plot, we show
                 the intrinsic upper limits on the lightest Higgs mass
                 for two values (2 and 16) of $\tan\beta=v_2/v_1$.

\item[Fig. 18] The cumulative probability distribution calculated from
the
                $\chi^2$ function in the SM shown in fig.~4, obtained
after
                 integrating
                appropriately over $m_t$, including the direct
measurements from
                 CDF   and D0  \cite{top}. This may be used to estimate
                the relative probabilities of different Higgs mass
ranges in the
                 SM and the MSSM, as discussed in the text \cite{efl} .

\item[Fig. 19] The expected significance for a Standard Model Higgs
boson in the ATLAS and
CMS experiments at the LHC \cite{atlas},  \cite{cms}. The higher-mass
range is also
accessible  up to about 1 TeV.

\item[Fig. 20] Capability of ATLAS$^{51}$     and CMS$^{52}$    to
explore the MSSM Higgs
sector. The regions with shaded edges can be explored with the channels
indicated. Also shown
is the region accessible to LEP2. Between LHC and LEP2, essentially the
entire plane is
covered.

\item[Fig. 21] Cross-sections for squark and gluino production at the
LHC.

\item[Fig. 22] Expected missing transverse energy signal for squarks and
gluinos at the LHC,
compared with the Standard Model and experimental backgrounds in  ATLAS
 \cite{atlas}.

\end{itemize}

\end{document}